\documentclass[12pt,preprint]{aastex}
\usepackage{epsf}

\newcommand{\simgt}{\lower.5ex\hbox{$\; \buildrel > \over \sim \;$}}
\newcommand{\simlt}{\lower.5ex\hbox{$\; \buildrel < \over \sim \;$}}
\newcommand{\calD}{{\cal D}}
\begin{document}
\title{Power Spectrum Analysis of the 2dF QSO Sample Revisited}
\author{Kazuhiro Yamamoto}
\affil{Graduate School of Science, Hiroshima University, 
Higashi-Hiroshima, 735-8526, Japan}
\email{kazuhiro@hiroshima-u.ac.jp}

\begin{abstract}
We revisit the power spectrum analysis of the complete sample of the two 
degree field (2dF) QSO redshift (2QZ) survey, as a complementary test of 
the work by Outram et al. (2003). A power spectrum consistent with that 
of the 2QZ group is obtained.
Differently from their approach, fitting of the power spectrum is 
investigated incorporating the nonlinear effects, the geometric 
distortion and the light-cone effect. 
It is shown that the QSO power spectrum is consistent with 
the $\Lambda$ cold dark matter (CDM) model with the matter density 
parameter $\Omega_m=0.2\sim0.5$. Our constraint on the density 
parameter is rather weaker than that of the 2QZ group. 
We also show that the constraint slightly depends on the equation 
of state parameter $w$ of the dark energy. The constraint on $w$ 
from the QSO power spectrum is demonstrated, though it is 
not very tight.
\end{abstract}




\section{Introduction}

The 2dF QSO redshift (2QZ) survey has established that the QSO 
sample is a useful probe of cosmological models as a tracer of 
the large scale distribution of mass (Croom, et~al. 2001;, Hoyle, 
et~al. 2002). 
In general, constraints on cosmological parameters from
QSO sample are not very tight. However, the cosmological 
parameters estimated from the QSO sample have a unique 
implication for cosmology (Outram, et~al. 2003; Yamamoto 2003a). 
For example, the cosmological principle can be tested by 
comparing with the result from other observations such as
galaxy redshift survey and cosmic microwave background 
anisotropies.
The pioneering work on the QSO power spectrum analysis was
done by Hoyle et~al. (2002) with the 2QZ 10000 catalogue.
Recently Outram et~al. have reported the result of the similar 
analysis with the final 2QZ catalogue 
containing 22652 QSOs (2003). They have shown that the QSO 
power spectrum is consistent with the Hubble Volume
$\Lambda$CDM simulation. Furthermore, by fitting the
power spectrum with the $\Lambda$CDM model within linear 
theory of density perturbations, they obtained a constraint 
on the cosmological density parameters. 

In the modeling of the QSO power spectrum in Outram et~al (2003), 
however, the light-cone effect (Matarrese et~al. 1997; Matsubara, Suto 
\& Szapdi 1997; Yamamoto \& Suto 1999), the geometric distortion 
(Ballinger, Peacock \& Heavens 1996; Matsubara \& Suto 1996) and 
the nonlinear effects (Mo, Jing \& B$\ddot{\rm o}$rner 1997; 
Magira, Jing \& Suto 2000) are not taken into account. 
The neglect of these effects might fail to estimate the 
correct cosmological parameters. To test this point, we 
revisit the power spectrum analysis of the 2QZ sample. 
We have independently performed the power spectrum analysis  
of clustering with the complete 2QZ sample. Then we fit the 
2QZ power spectrum with theoretical template 
incorporating the effects, which are not considered 
in the work by Outram et al. (2003). 
The methodology in the present paper is almost same 
as that in the reference (Yamamoto 2002), 
in which the fitting of the 2dF QSO power spectrum from the 
10000 catalogue was investigated using an analytic approach.
Thus the primary purpose of the present paper is to test the 
robustness of the result by Outram et al. (2003) for 
independent determination of the power spectrum and for more 
careful modeling of the theoretical power spectrum, including 
the nonlinear effects, the geometric distortion and the
light-cone effect. 

On the other hand, Calvao et al. (2002) claimed that the equation 
of state of the dark energy $w$ might be constrained from the 
2dF QSO sample. Due to the geometric distortion effect,
the QSO redshift-space power spectrum may depend on $w$ even 
if the original matter power spectrum (or the transfer function) 
does not depend on $w$ (Yamamoto 2003b).
The strategy in the present paper is not the one proposed by 
Calvao et al. (2002), however, we check a constraint on 
$w$ by considering how the estimated density parameters 
depends on $w$ by the fitting of the power spectrum.
The second purpose of this paper is to test the equation 
of state of the dark energy $w$ using the QSO power spectrum. 
This paper is organized as follows: 
In section 2, we describe our power spectrum analysis.
In section 3, our theoretical modeling of the QSO power 
spectrum is explained. In section 4, constraint on the 
density parameters 
is discussed by fitting the 2QZ power spectrum.
Section 5 is devoted to summary and conclusions.
Throughout this paper we use the unit in which the 
light velocity equals $1$. 

\section{The power spectrum analysis}
\def\bfs{{\bf s}}
\def\bfk{{\bf k}}
In our power spectrum analysis, we use the complete sample of the 
full 2QZ survey, which is publicly available 
${\rm (http://www.2dfquasar.org/Spec\_Cat/)}$. 
The 2QZ survey covers two area of $5\times75$ deg${}^2$, 
one in the South Galactic Cap (SGC) and the other 
in the North Galactic Cap (NGC), respectively, in the range of redshift 
less than 3.
The survey area is defined by the equatorial coordinates
from $\alpha=21^{\rm h}40$ to $\alpha=3^{\rm h}15$ and 
$-32.5^{\rm o}\le\delta\le-27.5^{\rm o}$ in the SGC, 
and $9^{\rm h}50\le\alpha\le14^{\rm h}50$ and 
$-2.5^{\rm o}\le\delta\le2.5^{\rm o}$ in the NGC, 
respectively. 
The survey area of the NGC is jagged and we select a
simple rectangle area in our power spectrum analysis.
Then we use 10713 and 8443 QSOs in the SGC and the NGC, respectively,
in the range of redshift $0.2\le z\le 2.2$ \footnote{
The 2QZ group used the QSOs
in the range of the redshift $0.3\le z\le 2.2$, which 
is slightly different from our choice. 
This difference does not alter our result.}, 
incorporating the hole information publicly available.

We describe the estimator of the power spectrum adopted here.
Three dimensional map is constructed by introducing the 
distance 
\begin{eqnarray}
  s(z)=\int_0^z{dz'\over H_0\sqrt{0.3(1+z')^3+0.7}},
\label{defsz}
\end{eqnarray}
where $s$ is the comoving distance of the $\Lambda$ cold dark matter 
model with the density parameter $0.3$. 
We denote the density field by $n(\bfs)$ and the mean number density by 
$\bar n(\bfs)$, where $\bfs=s\gamma$ with $\gamma$ specifying the direction. 
Introducing a random synthetic density field $n_{\rm s}(\bfs)$,
which has mean number density $1/\alpha$ times that of 
$n(\bfs)$, we define the Fourier coefficient
\begin{eqnarray}
  {\cal F(\bfk)}=A^{-1/2}\int d\bfs 
  [n(\bfs)-\alpha n_{\rm s}(\bfs)] e^{i{\bfk}\cdot\bfs}
\label{fouk}
\end{eqnarray}
with $A=\int d\bfs \bar n(\bfs)^2$. 
The estimator of the power spectrum is defined
\begin{eqnarray}
  P(k)={1\over V_k}\int_{V_k} d\bfk |{\cal F}(\bfk)|^2,
\end{eqnarray}
where $V_k$ is the volume of a thin shell in the $\bfk$-space
with the radius $k$. In the case $\alpha\ll1$, the 
variance of the power spectrum is 
\begin{eqnarray}
  \Delta P(k)^2={2(2\pi)^3 \over AV_k}.
\label{DP}
\end{eqnarray}
Note that we have not used the optimal weighting scheme by 
setting the optimal weight factor being constant (Feldman,
Kaiser and Peacock 1994, Tegmark et al. 1998, 
Yamamoto 2003b). This choice does
not alter the result of the QSO power spectrum analysis 
because the QSO is sparse and $\bar n P(k)<1$. 
Instead of equation (\ref{fouk}), the discrete density 
field can be rephrased as 
\begin{eqnarray}
  {\cal F(\bfk)}=A^{-1/2}
  [\sum_{i} n(\bfs_i)e^{i\bfk\cdot\bfs_i}-\alpha \sum_{j} n_{\rm s}(\bfs_j)
  e^{i\bfk\cdot\bfs_j}] ,
\label{foukdis}
\end{eqnarray}
where $\bfs_i$ and $\bfs_j$ are the position of the $i$-th 
QSOs and the $j$-th random objects. 

Figure 1 plots the power spectrum (filled squares), which is 
obtained by combining results from the SGC and NGC data sets. 
Fig.1 shows a good agreement with the result (open squares) 
by Outram et~al.~(2003). 
However, the error bar of our power spectrum is larger than
that of 2QZ group. This can originate from the difference of 
the error estimator adopted, which we describe in more details 
below.
The solid curve in Fig.1 is the theoretical curve of the 
$\Lambda$ cold dark matter (CDM) model with the cosmological 
parameter, $\Omega_m=0.28$, $\Omega_b=0.045$, $h=0.7$, 
$\sigma_8=0.9$ and $n=1$, motivated by the WMAP result 
(Spergel et~al. 2003, see also next section). 

Finally in this section, we compare the error estimator of the 
power spectrum adopted in the present paper with that of the 
2QZ group. There are two differences. First, they used the 
equation 2.3.2 in paper by Feldman et~al. (1994). In the limit
that $\bar n$ is a constant, there is the discrepancy of the 
factor $2$, if $V_k$, the volume of a thin shell in the 
$\bf k$-space, is same. In our analysis we used 
$V_k=4\pi k^2 \Delta k$, where $\Delta k$ is the width
of the $k$ bin. Our error estimator is consistent with that
formulated by Tegmark et~al. (1998; see equation 62 in 
their paper). Second, the 2QZ group adopted the different
method for $V_k$, estimated by $V_k=N_k(\Delta k)^3$ with 
$N_k$  the number of independent modes in the $k$-shell 
and $(\Delta k)^3$ the volume of one $k$-mode (for details 
see also Hoyle, et al. 2002; Hoyle 2000). 

\section{Modeling the theoretical power spectrum}
The author considered a constraint on the density parameters, 
$\Omega_m$ and $\Omega_b$, by fitting the 2dF QSO power spectrum 
found by Hoyle et~al. (2002), in a previous investigation (Yamamoto 2002).
We here adopt the similar methodology to apply it to the 2QZ power 
spectrum obtained in the previous section. 
The important improvement of the present work is the inclusion of 
nonlinear effects in modeling the theoretical power spectrum and 
a systematic uncertainty in measuring the redshift in the 2QZ survey, 
as well as the inclusion of an arbitrary equation of state parameter 
$w$ of the dark energy. In this section we briefly explain the modeling,
following the same notation in the previous paper. 

We restrict ourselves to a spatially flat FRW universe and follow
the quintessential cold dark matter (QCDM) model. The effective
equation of state of the dark energy, $w$, can be a function of
redshift, however, we consider the QCDM model with a constant 
equation of state for simplicity (Wang et~al. 2000). 
In this case the relation between the comoving distance and the redshift is 
\begin{eqnarray}
&&r(z)={1\over H_0}\int_0^z {dz'\over \sqrt{\Omega_m(1+z')^3+(1-\Omega_m)(1+z')^{3(1+w)}}},
\label{defrz}
\end{eqnarray}
where $H_0=100h$km/s/Mpc is the Hubble constant. 
Wang \& Steinhardt (1998) have given a useful approximate formula 
for the linear growth index in the QCDM model, which we adopt in 
the present work. 
For nonlinear modeling of the mass density perturbation, we adopt 
the simple fitting formulae for the nonlinear mass perturbation 
power spectrum presented by Ma et al. (1999) for the QCDM model,
combined with the transfer function by Eisenstein \& Hu (1998), 
which is robust even when the baryon fraction is large.

To incorporate the nonlinear (Finger-of-God) effect due to the 
random motion of objects, we consider the power spectrum multiplied
by the damping factor $D[q_{||}\sigma_{\rm P}]$, where $\sigma_{\rm P}$ 
is the pairwise velocity dispersion and $q_{||}$ is the comoving wave 
number parallel to the line-of-sight of an observer in real space.  
We adopt the expression assuming an exponential 
distribution function for the pairwise velocity, and the 
corresponding damping factor is (Mo, et~al. 1997;
Magira, et~al. 2000)
\begin{eqnarray}
  D[q_{||}\sigma_{\rm P}]={1\over 1+ q_{||}^2\sigma_{\rm P}^2/2}.
\label{Dsigmap}
\end{eqnarray}
The pairwise velocity dispersion is the function of redshift $z$. 
We computed $\sigma_{\rm P}^2$ 
by the approximate formula for the mean squire velocity dispersion at the 
large separation determined through the cosmic energy equation 
(Mo, et~al.~1997; Magira, et~al.~2000; Suto et~al. 2000). 

On the other hand, as described in the paper by Outram et~al.
(2003), systematic uncertainty in measurement of the redshift 
causes an apparent velocity dispersion. 
We incorporate the uncertainty in redshift-measurement
$\delta z$, based on the following consideration.
We write the density fluctuation field $\delta(\bfs)$
in a Fourier expansion form as 
\begin{eqnarray}
  \delta(\bfs)=\sum_\bfk \delta_\bfk e^{i\bfk\cdot (\bfs+\Delta\bfs)},
\end{eqnarray}
where $\Delta\bfs$ represents the error in measuring the position,
which is related with $\delta z$ by $|\Delta \bfs|(=\Delta s)=\delta z(ds/dz)$. 
Then an ensemble average of $\delta(\bfs)\delta(\bfs')$ can be
written as
\begin{eqnarray}
  \left\langle\delta(\bfs)\delta(\bfs')\right\rangle
  =\sum_{\bfk}  \langle|\delta_\bfk|^2\rangle e^{i\bfk\cdot(\bfs-\bfs')}
  \left\langle e^{i\bfk\cdot(\Delta\bfs-\Delta\bfs')}\right\rangle,
\end{eqnarray}
where we have assumed that $\delta_\bfk$ and $\Delta \bfs$ are
independent probability variables and 
$\langle\delta_\bfk\delta_{\bfk'}\rangle=\langle|\delta_\bfk|^2\rangle
  \delta^{(3)}(\bfk-\bfk')$. We further assume that
the angular coordinates are well determined and $\Delta \bfs=\gamma\Delta s$.
In this case, denoting the wave number of the line-of-sight 
direction by $k_{||}$, we have
\begin{eqnarray}
   \left\langle e^{i\bfk\cdot(\Delta\bfs-\Delta\bfs')}\right\rangle
   =e^{-k_{||}^2\left\langle(\Delta s-\Delta s')^2/2\right\rangle}
   =e^{-k_{||}^2\left\langle\Delta s^2\right\rangle},
\end{eqnarray}
where we have assumed that $\Delta s$ and $\Delta s'$ are 
independent Gaussian probability variables with the variance 
$\langle\Delta s^2\rangle$. Therefore the damping factor 
due to the error in redshift-measurement is written 
\begin{eqnarray}
  \calD[{\delta z}]= \exp\left[-k_{||}^2 \left({ds\over dz}\right)^2 
  \langle\delta z^2\rangle \right],
\label{delz}
\end{eqnarray}
where $\langle\delta z^2\rangle$ is the variance of the
error in measuring the redshift.
In the present paper we adopt $\delta z=0.0027z$ following Croom et al. 
(2003). Figure 2 shows the damping factor $D[k_{||}\sigma_{\rm P}]$ (solid 
curves) in the $\Lambda$CDM model\footnote{For definiteness, $D[k_{||}\sigma_{\rm P}]$ denotes 
the damping factor $D[q_{||}\sigma_{\rm P}]$ with replacing
$q_{||}$ with $(ds/dr)k_{||}$.},
as function of the wave number $k_{||}$ with the redshift fixed as 
$z=0.5$, $z=1$ and $z=1.5$ for three curves from top to bottom. The dashed 
curves show the damping factor $D[k_{||}\sigma_{\rm P}]$ multiplied by 
$\calD[{\delta z}]$. Because we are considering the angular averaged power 
spectrum, and this figure does not exactly express the damping factor of 
our power spectrum. However, this figure indicates that the damping factor 
have a substantial effect on the power spectrum shape. 
Thus the uncertainty in redshift-measurement can be an important factor 
in the power spectrum analysis. 

Concerning the modeling of the bias, we assume the scale independent model. 
Following the previous work (Yamamoto 2002), 
we consider the model $b(z)=b_0/D_1(z)^\nu$, 
where $b_0$ and $\nu$ are the constant parameters and $D_1(z)$ is the 
linear growth rate normalized as $D_1(0)=1$.
We determine $b_0$ to minimize $\chi^2$, which we define in the next 
section. In the present paper, we show the result adopting $\nu=1$.
Our result slightly depends on $\nu$, however, plausible 
alternation of $\nu$ does not alter our conclusion qualitatively.
(see also Yamamoto 2002).
We consistently use the number density per unit redshift per unit solid 
angle $dN/dz$ from the catalogue. 

It will be useful to show how the theoretical power spectrum depends on 
the equation of state $w$ of the dark energy. Curves in Figure 3 are
theoretical power spectra with the various equation of state 
$w=-1$(solid curve), $w=-2/3$(long dashed curve) and $w=-1/3$(dashed curve). 
The other cosmological parameter is fixed as $\Omega_m=0.28$, $\Omega_b=0.1$,
$h=0.7$, $\sigma_8=0.9$ and $n=1$. 
The difference of the three curves comes from the geometric distortion effect:
The comoving distance in real space is given by equation (\ref{defrz}), while
the power spectrum is measured in redshift space with the radial coordinate 
$s(z)$ defined by equation (\ref{defsz}). Therefore the power spectrum 
in redshift space is distorted compared with that in real space.
Denoting the wave number in real space by $q$, the geometric distortion 
is described by the scaling of wave numbers $q\rightarrow k/c$, 
where $c$ is the ratio $dr/ds$ and $r/s$ for the component 
parallel and perpendicular to the line-of-sight direction,
respectively (for details see e.g., Ballinger et~al. 1996, 
Yamamoto 2002; 2003b). 
For the model with $w>-1$, the factor $c$ becomes less than unity. 
This shifts the power spectrum from right (large $k$) to left (small $k$) 
for the model with $w>-1$, as is clearly shown in Fig. 3. Thus $w$ can be 
measured from the redshift power spectrum due to the scaling effect of the 
geometric distortion. Such a test using the scaling effect is distinguished 
from the Alcock-Paczynski test which relies on the anisotropy of clustering 
in redshift space (e.g., Ryden 1995; Alcock \& Paczynski 1979).

\section{Discussions}
In this section we compare our observational power spectrum from the
2QZ catalogue with the theoretical model to constrain cosmological 
parameters. We introduce
\begin{eqnarray}
  \chi^2=\sum_{i=1}^N{[P^{\rm th}(k_i)-P^{\rm ob}(k_i)]^2\over \Delta P(k_i)^2}, 
\end{eqnarray}
where $P^{\rm th}(k_i)$ is the value of the power spectrum from the 
theoretical model at the wave number $k_i$, $P^{\rm ob}(k_i)$ is the observational 
data and $\Delta P(k_i)$ is the variance of errors, for which we use the $(N=)20$ 
data points obtained in Fig.~1.
Following the analysis by Outram et~al. (2003), we consider the likelihood 
function $\propto e^{-\chi^2/2}$. 
Figure 4 shows contours of the likelihood function with $\chi^2$ found when different values of
the cosmological density parameters, $\Omega_m$ and $\Omega_b/\Omega_m$
are used in the theoretical modeling. Each panel corresponds to
constraints for the difference model of $P^{\rm th}(k)$: 
The panels (a), (c) and (d) assume $w=-1$, $-2/3$ and $-1/3$, respectively.  
The panel (b) is same as (a), but switched off the damping factor due to the
error in the redshift-measurement by setting $\delta z=0$ in equation (\ref{delz}).
In Fig.~4, we fix the other parameters $h=0.7$, $n=1$ and $\sigma_8=0.9$. 
The curves are contours of confidence $65\%$, $95\%$ and $99\%$ on 
the plane. 

First, the panel (b) in Fig.~4 shows the effect of the error in the 
redshift-measurement in modeling the power spectrum. 
The peak value of (a) is located  at $\Omega_m=0.27$, while that of (b) 
is located at $\Omega_m=0.24$. Thus the effect of $\delta z$ can be of 
importance for determining $\Omega_m$, precisely. 
Second, Fig.~4 demonstrates the 
constraint on the density parameters depends on the equation of state 
parameter $w$ of the dark energy. It shows that the preferable value of 
$\Omega_m$ and $\Omega_b$ increases as $w$ becomes large. 
Figure 5 shows the contour of the likelihood function 
with $\chi^2$ found when different values of $\Omega_m$ and $w$ are 
used in the 
modeling the power spectrum, where we fixed $\Omega_b=0.045$,
motivated from the WMAP result (Spergel et~al. 2003). 
It is shown that the density parameters in the range $0.2\simlt 
\Omega_m\simlt 0.4$ is preferable, which is not very sensitive 
to the value of $w$. Therefore the constraint on $w$ is not 
very tight, and $w\simgt -0.2 ~(-0.1)$ is only excluded 
at the one (two) sigma level. 

\section{Conclusion}

In summary we have revisited the power spectrum analysis of the complete 
2dF QSO sample, as a complementary test of the work by Outram et~al. 
(2003). Our analysis has reproduced a power spectrum, which is consistent 
with the 2QZ group. We have investigated the fitting of the power spectrum 
including the light-cone effect, the geometric distortion and the 
nonlinear effects.
It is shown that the QSO spatial power spectrum 
is consistent with the $\Lambda$CDM model with $\Omega_m=0.2\sim 0.5$,
which is effectively consistent with the 2QZ group. However, our
constraint is weaker than that of the 2QZ group. This will be
traced back to the difference of the error estimation of the 
power spectrum. 
In the present paper, we have emphasized the importance of 
the error in measurement of the redshift $\delta z$ because 
it may have influence in estimating the density parameters. 
We have also investigated the effect of the equation of state of the dark 
energy $w$.  For models with large value of $w>-1$, the preferable
value of the density parameters becomes large. When we fix the
baryon density parameter as that found by the WMAP team,
the matter density parameter $0.2\simlt\Omega_m\simlt 0.4$
is preferable. No tight constraint on $w$ is obtained,
$w\simgt-0.2 ~(-0.1)$ is only excluded at the one (two) sigma level. 
In general QSO sample is sparse and the shot noise contribution is 
substantial. Therefore the constraint from it is not very tight. 
However, such a power spectrum analysis will provide more useful 
constraint on $w$ when applied to the SDSS luminous red galaxy 
sample (Yamamoto 2003b) and next generation redshift survey 
such as the KAOS project (Seo and Eisenstein 2003).

\begin{acknowledgments}
The author thanks anonymous referee for useful comments on
earlier version of the manuscript, which helped improve it. 
This work is supported in part by Grant-in-Aid for
Scientific research of Japanese Ministry of Education, Culture, Sports, 
Science and Technology, No.15740155.
\end{acknowledgments}

\clearpage

\begin{figure}
\begin{center}
    \leavevmode
    \epsfxsize=16cm
    \epsfbox[20 150 600 720]{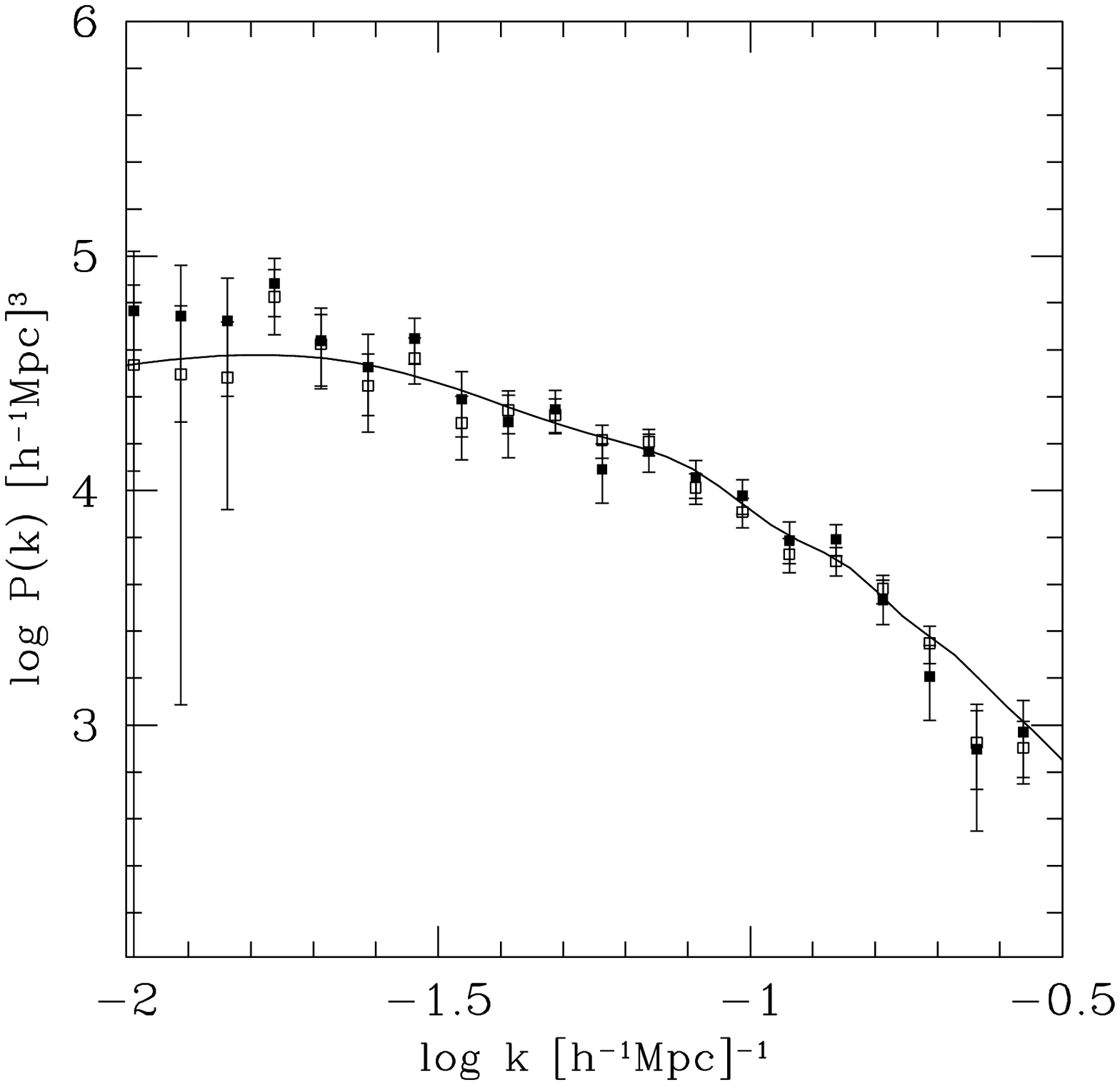}
\end{center}
\caption{
Power spectrum from the 2dF QSO sample. 
The filled square is the result of our analysis,
while the open square is from  Outram et al. (2003).
The solid curve is the theoretical curve of the $\Lambda$CDM 
model with the cosmological parameter, $\Omega_m=0.28$, 
$\Omega_b=0.045$, $h=0.7$, $\sigma_8=0.9$ and $n=1$.
(see also section 3)}
\label{figurea}
\end{figure}
\begin{figure}
\begin{center}
    \leavevmode
    \epsfxsize=16cm
    \epsfbox[20 150 600 720]{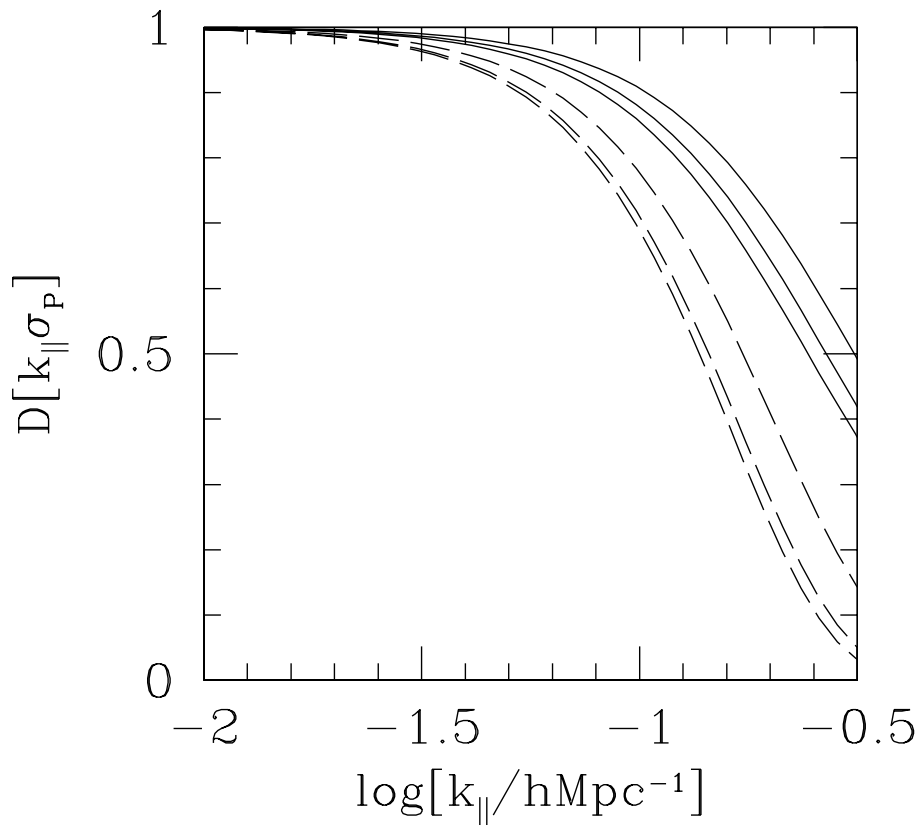}
\end{center}
\caption{Damping factor $D[k_{||}\sigma_P(z)]$ (solid curves) 
as function of the wave number $k_{||}$ with the redshift fixed 
$z=0.5$, $z=1.0$ and $z=1.5$ from top to bottom. 
The dashed curves show the damping factor multiplied by $\calD[{\delta z}]$.
We assumed the same theoretical model as that in Fig.~1.
}
\label{figureb}
\end{figure}
\begin{figure}
\begin{center}
    \leavevmode
    \epsfxsize=16cm
    \epsfbox[20 150 600 720]{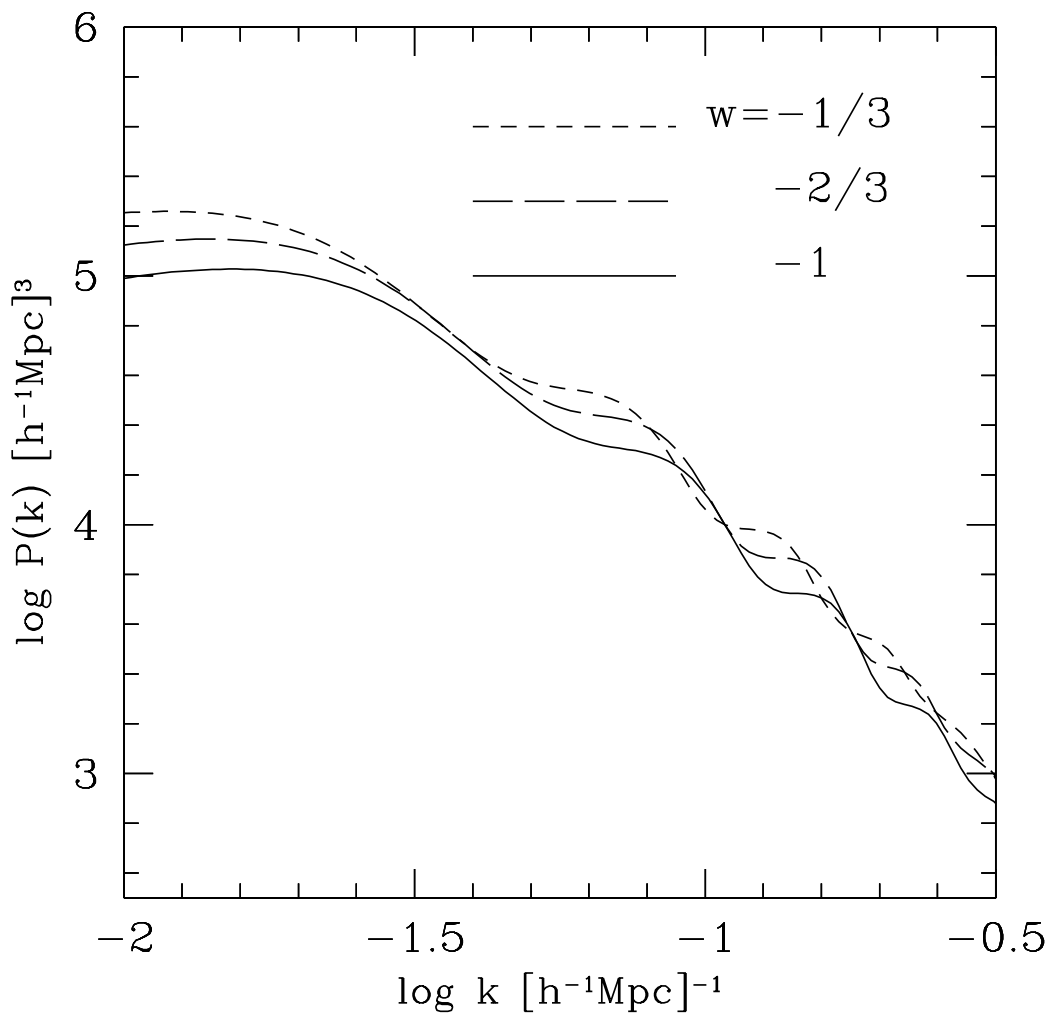}
\end{center}
\caption{
Theoretical power spectra with $w=-1$ (solid curve),
$w=-2/3$ (long dashed curve) and $w=-1/3$ (dashed curve).
The other cosmological parameter is fixed as $\Omega_m=0.28$, 
$\Omega_b=0.1$, $h=0.7$, $\sigma_8=0.9$ and $n=1$. The amplitude
is normalized as $P(k)=10^{4}(h^{-1}{\rm Mpc})^3$ at $k=0.1h{\rm Mpc}^{-1}$. 
Note that the model of large baryon fraction is adopted 
to emphasize the scaling effect of the geometric distortion.}
\label{figuref}
\end{figure}
\begin{figure}
\begin{center}
    \leavevmode
    \epsfxsize=16cm
    \epsfbox[20 150 600 720]{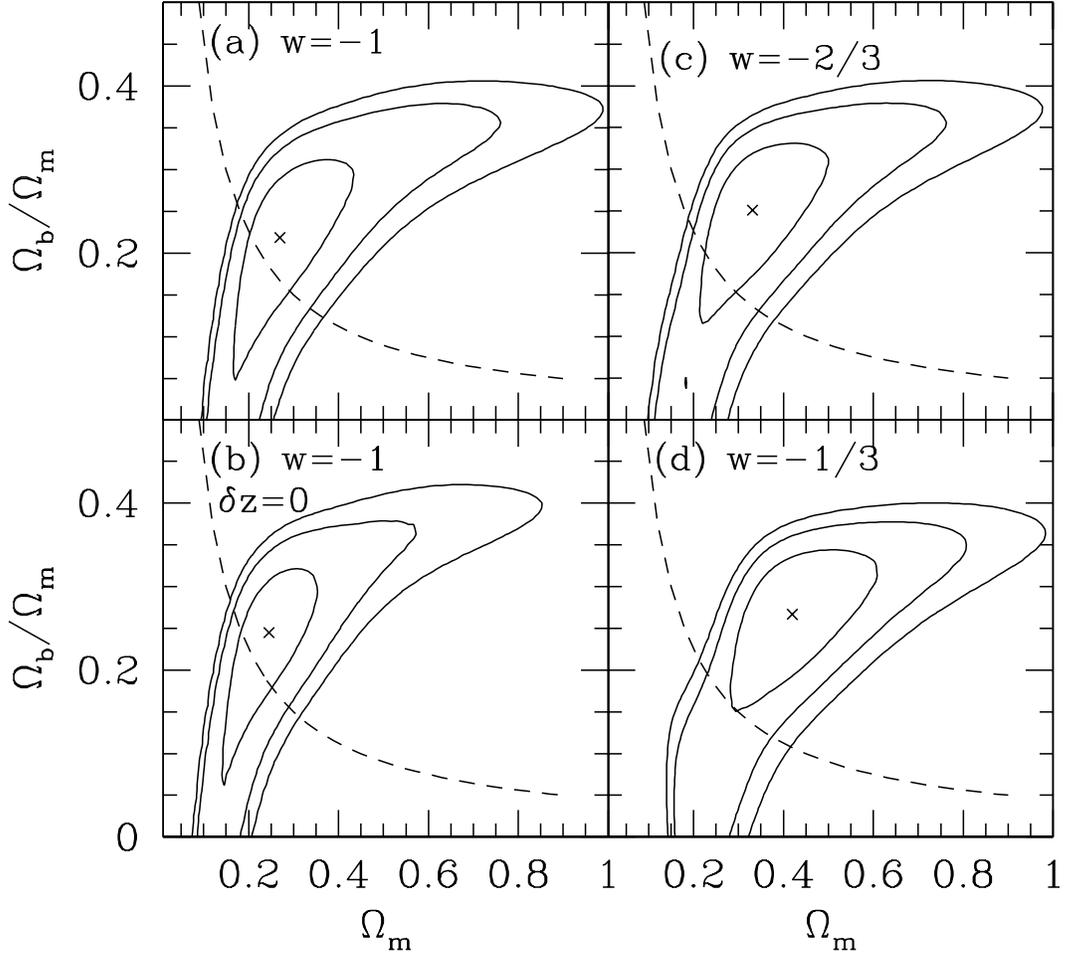}
\end{center}
\caption{Contours of the likelihood function of our QSO power spectrum.
The panel (a), (c) and (d) assumes $w=-1$, $-2/3$ and $-1/3$, respectively,
in which we adopt $h=0.7$, $n=1$ and $\sigma_8=0.9$. 
The panel (b) is same as (a)
but neglected the damping factor due to the error in redshift-measurement.
In each panel, contours are confidence of $65\%$, $95\%$ and $99\%$.
The cross point shows the best fitted parameter:
$(\Omega_m,\Omega_b)=(0.27,0.06)$, $(0.24,0.06)$,
$(0.33,0.08)$, $(0.41,0.11)$, from (a) to (d), respectively.
The dashed curve shows $\Omega_b=0.045$ for comparison. 
}
\label{figurec}
\end{figure}
\begin{figure}
\begin{center}
    \leavevmode
    \epsfxsize=16cm
    \epsfbox[20 150 600 720]{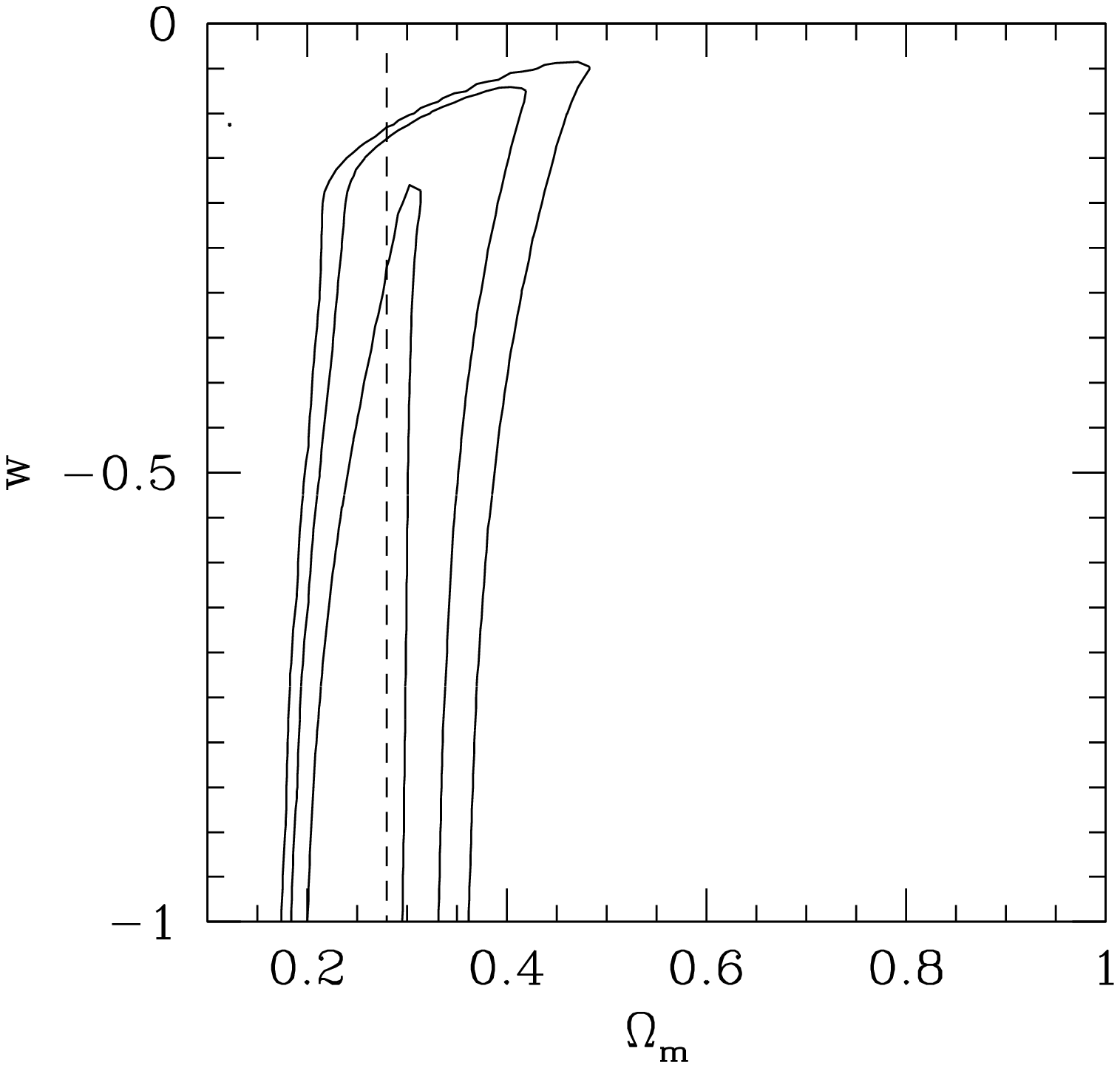}
\end{center}
\caption{Contours of the likelihood function on the $\Omega_m$ and $w$ plane.
Contours are confidence of $65\%$, $95\%$ and $99\%$.
Here we assume  $\Omega_b=0.045$, $h=0.7$, $n=1$ and $\sigma_8=0.9$.
The dashed line is $\Omega_m=0.28$.
}
\label{figured}
\end{figure}


\newpage
\begin{thebibliography}{99}
\bibitem[]{AP}
  Alcock, C., \& Paczynski, B. 1979, Nature, 281, 358
\bibitem[]{BPH}
  Ballinger, W. E., Peacock, J. A., \& Heavens, A. F. 1996, MNRAS, 282, 877
\bibitem[]{Calvao}
  Calvao, M. O., de Mello Neto, J. R. T., \& Waga, I.  2002, Phys. Rev. Lett. 88, 091302
\bibitem[]{Croomb}
  Croom, S. M., Shanks, T., Boyle, B. J., Smith, R.J., Miller, L., Loaring, N., 
  \& Hoyle, F. 2001, MNRAS, 325, 483
\bibitem[]{Croomc}
  Croom, S. M., Smith, R. J., Boyle, B. J., Shanks, T., 
  Miller, L., Outram, P. J., \& Loaring, N. S. 2003, preprint
\bibitem[]{EH}
  Eisenstein, D. J., \& Hu, W. 1998, ApJ, 496, 605
\bibitem{FKP}
  Feldman, H. A., Kaiser, N., \& Peacock, J. A. 1994, ApJ, 426, 23 
\bibitem[]{Hoyle2000}
  Hoyle, F., 2000, PhD thesis, Univ. Durham
\bibitem[]{Hoyle}
  Hoyle, F., Outram, P.J., Shanks, T., Croom, S.M., Boyle, B.J., Loaring, N., Miller, L., 
  \& Smith, R.J. 2002, MNRAS, 329, 336
\bibitem[]{Ma}
  Ma, C.-P., Calldwell, R. R., Bode, P., \& Wang, L. 1999, ApJ, 521, L1
\bibitem[]{Mat}
  Matarrese, S., Coles, P., Lucchin, F., \& Moscardini, L. 1997, MNRAS, 286, 115
\bibitem[]{MS}
  Matsubara, T., \& Suto, Y. 1996, ApJ, 470, L1
\bibitem[]{MSS}
  Matsubara, T., Suto, Y., \& Szapdi, I. 1997, ApJ, 491, L1
\bibitem[]{MJS}
  Magira, H., Jing, Y. P., \& Suto, Y. 2000, ApJ, 528, 30
\bibitem[]{MJB}
  Mo, J., Jing, Y. P., \& B$\ddot{\rm o}$rner, G. 1997, 
  MNRAS, 286, 979
\bibitem[]{NY}
  Nishioka, H., \& Yamamoto, K. 1999, ApJ, 520, 426
\bibitem[]{Outram}
  Outram, P.J., Hoyle, F., Shanks, T., Croom, S. M., 
  Boyle, B. J., Miller, L., Smith, R. J., 
  \& Myers, A. D. 2003, MNRAS, 342, 483
\bibitem[]{Ryden}
  Ryden, B. S. 1995, ApJ, 452, 25
\bibitem[]{SE}
  Seo, H-J, \& Eisenstein D. J. 2003, astro-ph/0307460
\bibitem[]{spergel}
  Spergel, D. N., et~al. 2003, ApJS, 148, 175
\bibitem[]{SMJMY}
  Suto, Y., Magira, H., Jing, Y. P., Matsubara, T., \& Yamamoto, K. 2000, 
  Prog. Theor. Phys. Suppl. 133, 183
\bibitem[]{THVS}
  Tegmark, M., Hamilton, A. J. S., Strauss, M. A., Vogeley, \& M. S., Szalay, A. S.
  1998, ApJ, 499, 555
\bibitem[]{WCOS}
  Wang, L., Caldwell, R. R., Ostriker, J. P., \& Steinhardt, P. J. 2000, ApJ, 530, 17
\bibitem[]{WS}
  Wang, L., \& Steinhardt, P. J. 1998, ApJ, 508, 483
\bibitem[]{YS}
  Yamamoto, K., \& Suto, Y. 1999, ApJ, 517, 1
\bibitem[]{Y2002}
  Yamamoto, K. 2002, MNRAS, 334, 958 
\bibitem[]{Y2003a}
  Yamamoto, K. 2003a, MNRAS, 341, 1199
\bibitem[]{Y2003b}
  Yamamoto, K. 2003b, ApJ, 595, 577
\end{thebibliography}
\end{document}